\documentstyle[amssymb,secnumtab,epsf]{fbssuppl}

\title{Front-form calculation of the deuteron EM form factors}

\author{J. Carbonell\thanks{e-mail: carbonel@isn.in2p3.fr}  \instnr{1},
        V.A. Karmanov\thanks{e-mail: karmanov@sci.lebedev.ru}\instnr{2}}

\instlist{Institut des  Sciences Nucl\'{e}aires, 
          53 av. des Martyrs, 38026 Grenoble, France
\and      Lebedev Physical Institute, 
          Leninsky Prospekt 53, 117924 Moscow, Russia}

\begin{document}

\maketitle

\begin{abstract}
The deuteron form factors are calculated in the framework of 
the relativistic nucleon-meson dynamics.  The relativistic effects 
change considerably the S- and D-waves of the deuteron, result in the 
dominating extra component in the deuteron wave function and generate 
the contact interaction between the nucleon, mesons and photon.  The 
prediction for the polarization observable $t_{20}$ is in good 
agreement with the recent data of TJNAF/CEBAF.
\end{abstract}

\section{Introduction}\label{intro}

We calculate the deuteron form factors in the framework of the 
explicitly covariant version of light-front dynamics (see for review 
\cite{cdkm}). In this approach one can use in full measure the 
knowledge of the nonrelativistic wave functions and incorporate 
selfconsistently the relativistic effects. We assume that the nucleons 
interact with each other by exchanging the same mesons that are 
incorporated in the $NN$ potential, however we don't use any 
nonrelativistic potential approximation.

\section{Wave function}\label{wf}

In the explicitly covariant formulation of light-front dynamics the 
wave function is defined in the space-time on the plane formally given 
by the  light-front  equation $\omega\cdot x =0$, where $\omega$ is the 
four-vector with $\omega^2=0$.  This restores the relativistic 
covariance in comparison to the standard light-front approach. The 
latter is obtained as a particular case for $\omega=(1,0,0,-1)$.  The 
wave functions depend on the orientation of the light front, as they 
should.  This dependence is covariantly parametrized in terms of the 
four-vector $\omega$.   

The explicit covariance allows to construct the general form of the 
light-front wave function for a system with given spin. The 
relativistic deuteron wave function on the light front contains six 
spin components, in contrast to two components -- S and D-waves -- in 
the nonrelativistic case. Its general form was found in \cite{karm81}.  
In our calculation we keep only three dominating components $f_1,f_2$ 
and $f_5$.

The relativistic deuteron wave function was calculated in 
\cite{ck-deut} by a perturbative way incorporating exact relativistic 
OBEP kernel found in light-front dynamics, and the nonrelativistic wave 
function as the starting point. For the kernel the set of six mesons 
($\pi,\eta,\rho,\omega,\delta$ and $\sigma$) and the parameters 
corresponding to the Bonn model \cite{bonn} were used.

In nonrelativistic region of $k$ the components $f_1$ and $f_2$ found 
by this way turn into the usual S- and D-waves, whereas other 
components become negligible. However, starting from $k\approx 0.5$ 
GeV/c the component $f_5$ dominates over all other components, 
including $f_1$ and $f_2$. 

The physical meaning of this dominating extra component has been 
clarified \cite{dkm95} by comparing, in $1/m$ approximation, the 
analytical expression for the amplitude of the deuteron 
electrodisintegration near threshold with the nonrelativistic one, 
including meson exchange currents. For the isovector transition, in the 
region, where so called pair term with the pion exchange dominates, 
this component (together with similar component in the scattering 
state) automatically incorporates 50\% of the contribution of the pair 
term and therefore dominates too. Another 50\% is given by the contact 
(instantaneous) interaction (see below).

\section{Electromagnetic vertex}\label{emv}

A peculiarity of the light-front dynamics is the presence of the so called 
instantaneous (or contact) interaction $NNB\gamma$, where $B$ is the 
meson $\pi,\rho$, etc. This is a trace of disappeared diagrams 
corresponding to vacuum fluctuations. The fermion fields in this 
interaction are sandwiched with the factor 
$\hat{\omega}=\omega_{\mu}\gamma^{\mu}$ that in the standard approach 
is the well known $\gamma^+$ (in addition to the standard factors of 
the nucleon-meson Lagrangian, like $\gamma_5,\gamma_{\mu}$, etc.).  
Therefore the full electromagnetic vertex which we use to calculate the 
deuteron form factors corresponds to the sum of the impulse 
approximation and of the contact interaction.  It partially takes into 
account the many-body currents.  For the contact interaction we take 
into account the sum over all six mesons contributing to the Bonn 
potential, with the parameters of the Bonn model \cite{bonn}.

\section{Finding form factors}\label{ffs}

The next step is to extract the deuteron form factors from the 
electromagnetic vertex.  The physical amplitudes 
should not depend on the orientation of the light-front plane.  However in 
practice, because of the incompatibility between the transformation properties 
of the approximate e.m. current and the wave function, the 
nonphysical $\omega$ dependence survives in the on-energy shell 
deuteron electromagnetic vertex.  Due to covariance, the general form 
of this dependence can be found explicitly and then separated from the 
physical form factors.  The explicit formulas extracting the physical 
form factors ${\cal F}_1, {\cal F}_2, {\cal G}_1$ from the nonphysical 
contributions were found in refs. \cite{ks9294}. 

The polarization observable $t_{20}$ measured in experiment is a 
combination of the form factors (see e.g. \cite{cdkm}). Besides 
momentum transfer it depends also explicitly on 
$\tan^2\frac{1}{2}\theta$.

\section{Numerical results and conclusions}\label{numer}

Our prediction \cite{cdkm,ck_98} for the polarization observable $t_{20}$ 
together with the most recent measurements are shown in figure 1.
The calculation is in close agreement with the 
existing experimental data in all the momentum range,
including the new results obtained at TJNAF/CEBAF \cite{kox}
which corresponds to $\theta=70^{\circ}$. 

\begin{figure}[htbp]
\epsfxsize=11cm\epsfysize=9cm
\begin{center}
\epsfbox{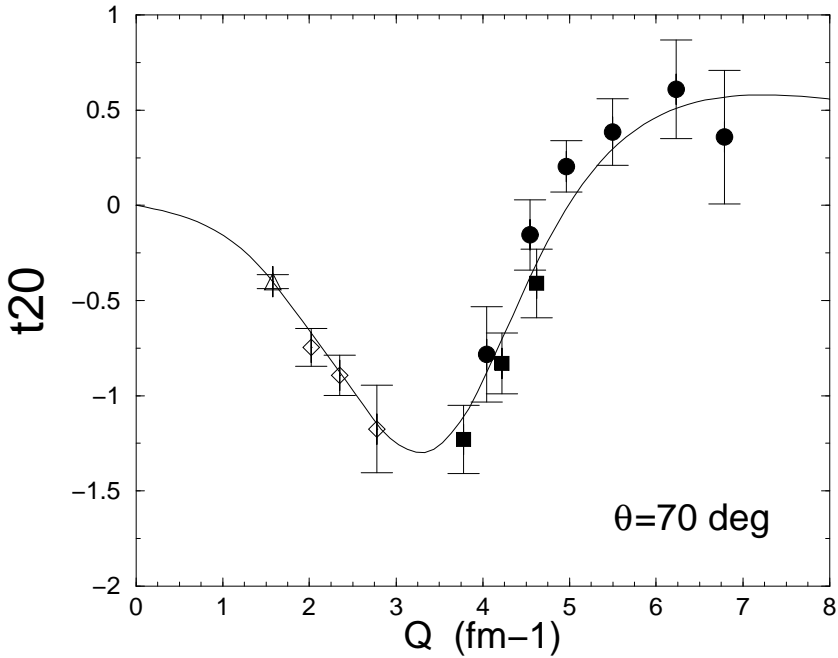}
\end{center}
\caption{The deuteron tensor polarization $t_{20}$ at $\theta=70^{\circ}$ 
compared to most recent experimental data: triangle from \cite{NIKHEF_96}, 
diamonds from \cite{NIKHEF_97}, squares from \cite{Bates_91}, circles from \cite{kox}}\label{fig1}
\end{figure}

This coincidence means that the deuteron structure at small distances 
is understood rather well. It is given by the relativistic 
nucleon-meson dynamics. In its turn, the relativistic nucleon-meson 
dynamics is well described theoretically, in the framework of the 
phenomenological nucleon-meson Lagrangian. The parameters of this 
Lagrangian are found phenomenologically. Their calculation in the 
framework of QCD is a separate problem. 

This allows us to make the general conclusion that: 
{\it in a large extent the relativistic nuclear physics can be developed 
independently of its derivation from QCD.}

However, these results can be improved in many respects.  An exact 
solution of the equation for the deuteron is indeed feasible. It 
implies the redefinition of the parameters of the NN kernel in the 
framework of the light-front equations.  This necessitates also a 
careful treatment of higher order contributions to the kernel. Finally, 
the calculation of electromagnetic observables should include higher 
Fock states ($NN\pi$, etc.) and the meson exchange currents 
corresponding to the direct interaction of the photon with the 
intermediate mesons.

\section*{Acknowledgement}
The authors are sincerely grateful to B. Desplanques and J.-F. Mathiot 
for many useful discussions, valuable remarks and helpful advices.

\end{document}